\documentclass[showpacs,preprintnumbers,
amsmath,amssymb,aps,prd]{revtex4}
\usepackage{amsfonts}
\usepackage{amsfonts,graphics,epsfig,subfigure}

\begin{document}

\title{Phase transition and heat engine efficiency of phantom AdS black holes}
\author{Jie-Xiong Mo, Shan-Quan Lan}
\affiliation{Institute of Theoretical Physics, Lingnan Normal University, Zhanjiang, 524048, Guangdong, China}

\begin{abstract}
 Phase transition and heat engine efficiency of phantom AdS black holes are investigated with peculiar properties found. In the non-extended phase space, we probe the possibility of $T-S$ criticality in both the canonical ensemble and grand-canonical ensemble. It is shown that no $T-S$ criticality exists for the phantom AdS black hole in the canonical ensemble, which is different from the RN-AdS black hole. Contrary to the canonical ensemble, no critical point can be found for neither phantom AdS black holes nor RN-AdS black hole in the grand-canonical ensemble. Moreover, we study the specific heat at constant electric potential. When the electric potential satisfies $A_0>1$, only phantom AdS black holes undergo phase transition in the grand-canonical ensemble. In the extended phase space, we show that there is no $P-V$ criticality for phantom AdS black holes, contrary to the case of the RN-AdS black hole. Furthermore, we define a new kind of heat engine via phantom AdS black holes. Comparing to RN-AdS black holes, phantom AdS black holes have a lower heat engine efficiency. However, the ratio $\eta/\eta_C$ of phantom AdS black hole is higher, thus increasing the possibility of approaching the Carnot limit. This observation is obviously of interest. The interesting results obtained in this paper may be attributed to the existence of phantom field whose energy density is negative.
\end{abstract}
  \maketitle

\section{Introduction}

    Ever since the discovery of the well-known Hawking-Page phase transition~\cite{Hawking2} which describes the phase transition between the Schwarzschild AdS black hole and the thermal AdS space, phase transition of AdS black holes has long been an interesting research topic. Chamblin et al. studied the phase transition of RN-AdS black holes \cite{Chamblin1,Chamblin2} and revealed the close relation between charged AdS black holes and the liquid-gas system. Recently, Kubiz\v{n}\'{a}k and Mann further enhanced this relation by probing the $P-V$ criticality \cite{Kubiznak} in the extended phase space. The extended phase space here refers to including the thermodynamic pressure and thermodynamic volume in the phase space \cite{Dolan2}. Interpreting the cosmological constant as a variable \cite{Caldarelli}-\cite{Cvetic}, thermodynamics especially phase transition of black holes in the extended phase space has received more and more attention. One can refer to the recent review \cite{Kubiznak2} and references therein.

Recent cosmological observations suggests the Universe appears to be expanding at an increasing rate. To explain this phenomenon, varieties of models have been proposed. One kind of candidates may be the modified gravity theories, such as $f(R)$ gravity. Another way is to introduce an effective field generating repulsive gravity. Among the latter approach, exotic fields represented by a distribution with negative energy density can be introduced to solve the problem of cosmic acceleration. Phantom fields serve as such candidates that explain the observational data of the cosmic acceleration quite well \cite{Dunkle,Hannenstadt}. So it is of great interest to investigate the physical properties of phantom field, especially from the context of gravity theories.

The first motivation of this paper is to present a unified phase transition picture of phantom AdS black holes, not only in the non-extended phase space but also in the extended phase space. For the sake of comparison, we choose the phantom AdS black hole solution proposed in Ref. \cite{Jardim} as our research subject. This solution was obtained while Einstein-Hilbert action with the cosmological constant is coupled with the Maxwell field or the phantom field. When $\eta=-1$, it represents phantom AdS black holes and it reduces to RN-AdS black holes when $\eta=1$. So it is very convenient to compare the results with those of RN-AdS black holes and disclose the effects of phantom fields.

Concerning the phase transition of phantom black holes, many efforts have been made. Ref. \cite{Jardim} obtained the basic quantities of phantom AdS black holes in the Einstein-Maxwell theory, such as the Hawking temperature, the entropy, the specific heat at constant charge and the free energy. Moreover, it pointed out the inconsistency between the classical analysis and geometrothermodynamics. Although it claimed that local and global stabilities have been established through the specific heat and the canonical and gran-canonical ensembles, the specific heat at constant electric potential (an important tool to probe the stability in the grand-canonical ensemble) was missed in their analysis. Ref. \cite{Quevedo} shows that the classical analysis and geometrothermodynamics can be compatible if one consider the cosmological constant as a thermodynamic variable. Instead, they introduced the term $Ldl$ into the first law, where $L$ is the intensive variable dual to $l$. This treatment is in accordance with the spirit of the extended phase space, although the definition of the physical quantities is varied. Moreover, their analysis was concentrated on the geometrothermodynamics, leaving the phase transition in the extended phase space (especially the existence of $P-V$ criticality) unexplored. Ref. \cite{Rodrigues1} studied the asymptotically flat black hole solutions of the Einstein-Maxwell-Dilaton theory (where the effect of dilaton field is included) and their thermodynamics. Their critical phenomena were investigated in \cite{Rodrigues2} and geometrothermodynamics was revisited in \cite{Quevedo2} with a novel type of phase transition found. As stated in the former paragraph, we aim at a unified phase transition picture of phantom AdS black holes.

Besides the phase transition, we are also interested in the heat engine utilizing phantom AdS black holes as working substance. The concept of holographic heat engine was creatively introduced in Ref. \cite{Johnson1}, allowing one to extract useful mechanical work from black holes. It was argued that the engine cycle represents a journal through a family of holographically dual large $N$ field theories \cite{Johnson1}, thus having interesting holographic implications. Subsequently, a lot of efforts have been devoted to this topic and varieties of interesting findings are emerging \cite{Johnson2}-\cite{Rosso}. It is worth mentioning that effects of quintessence dark energy on the heat engine was studied in Ref. \cite{mengxinhe}. One may wonder whether phantom dark energy exerts influence on the heat engine efficiency, just as the quintessence dark energy does. And this is partly why we carry out the work in this paper.

    The organization of this paper is as follows. We will present a short review of the thermodynamics of phantom AdS black holes in Sec.\ref{Sec2}. Phase transition of phantom AdS black holes in the non-extended phase space will be complemented in Sec.\ref {Sec3} while phase transition will be probed in the extended phase space in Sec.\ref {Sec4}. Moreover, we will study the heat engine efficiency of phantom AdS black holes in Sec.\ref{Sec5}. And the paper will be end with a brief conclusion presented in Sec.\ref {Sec6}.

\section{Review of thermodynamics of phantom AdS black holes}
\label{Sec2}

The action of Einstein theory with cosmological constant minimally coupled to the electromagnetic field reads \cite{Jardim}
\begin{equation}
S=\int d^4 x\sqrt{-g}\left(R+2\eta F_{\mu\nu}F^{\mu\nu}+2\Lambda\right)\label{1}\,,
\end{equation}
where $\Lambda$ denotes the cosmological constant and $\eta$ is a constant which indicates the nature of the electromagnetic field. For the Maxwell field, $\eta=1$ while $\eta=-1$ for the phantom field of spin $1$. Note that the energy density of this phantom field is negative.

The solution of the above action has been obtained as
\begin{eqnarray}
ds^{2}&=&f(r)dt^2-\frac{1}{f(r)}dr^2-r^2\left(d\theta^2+\sin^2\theta d\phi^2\right)\;, \nonumber\\
F&=&-\frac{q}{r^2}dr\wedge dt\;,\;f(r)=1-\frac{2M}{r}-\frac{\Lambda}{3}r^2+\eta\frac{q^2}{r^2}\;,\label{2}
\end{eqnarray}
where $M, q$ denote the mass of the black hole and the electric charge of the source respectively. When all the parameters vanish, this solution reduces to Minkowski spacetime. It was argued that it is asymptotically anti-de Sitter for $\Lambda<0$ and recovers the Reissner-Nordstr\"{o}m-AdS (RN-AdS) black hole when $\eta=1$ \cite{Jardim} .

Solving the equations $f(r)=0$, one can obtain two positive real roots for the case $\eta=1$ (corresponding to the event horizon and the internal horizon) and only one positive root for the case $\eta=-1$ (corresponding to the event horizon) \cite{Jardim} .

The mass, the Hawking temperature, the entropy and the electric potential was obtained in Ref. \cite{Jardim} as
\begin{eqnarray}
M&=&\frac{r_+}{2}\left(1-\frac{\Lambda}{3}r_+^2+\eta\frac{q^2}{r_+^2}\right)\label{3}\;,\\
T&=&\frac{1}{4\pi r_+}\left(1-\Lambda r_+^2-\eta\frac{q^2}{r_+^2}\right)\;.\label{4}\\
S&=&\frac{1}{4}A=\pi r_{+}^{2}\label{5}\;.\\
A_{0}&=&\frac{q}{r_{+}}\label{6}\;.
\end{eqnarray}

With the above physical quantities, the first law of the thermodynamics was presented as \cite{Jardim}
\begin{equation}
dM=TdS+\eta A_{0}dq \label{7}\;.
\end{equation}
For the case $\eta=-1$, the second term in the right hand side of the above equation changes its sign implying it contributes negative energy to the system.
Ref. \cite{Quevedo} argued that it is necessary to consider the cosmological constant as a variable and proposed the extended version of the first law as
\begin{equation}
dM=TdS+Ldl+\eta A_{0}dq \label{8}\;,
\end{equation}
where $l^2=-3/\Lambda$. $L$ is the intensive variable conjugate to $l$ related by $L=-(r_+/l)^3$ \cite{Quevedo}.
In this paper, we would like to probe the phase transition of phantom AdS black holes in both the non-extended phase space (where the cosmological constant is viewed as constant) and extended phase space ((where the cosmological constant is viewed as a thermodynamic variable).

\section{Phase transition of phantom AdS black holes in the non-extended phase space}
\label{Sec3}

The specific heat at constant $q$ was derived in Ref. \cite{Jardim} as
\begin{equation}
C_q=2S\frac{-\pi S+\Lambda S^2+\eta \pi^2q^2}{\pi S+\Lambda S^2-3\eta \pi^2q^2} \label{9}\;.
\end{equation}
And it was observed that only one phase transition point (also the divergent point of $C_q$) exists for the case $\eta=-1$.

Here, we continue to investigate the $T-S$ criticality, which has become a hot topic in recent years since the original work of Spallucci and Smailagic \cite{Spallucci}(where $T-S$ criticality was observed for RN-AdS black holes). One may wonder whether there exists similar phenomena for the phantom AdS black holes.
Utilizing Eqs. (\ref{4}) and (\ref{5}), one can obtain
\begin{eqnarray}
\left(\frac{\partial T}{\partial S}\right)_{q}&=&-\frac{\pi S+\Lambda S^2-3\eta \pi^2q^2}{8\pi^{3/2}S^{5/2}},\label{10}
\\
\left(\frac{\partial^2 T}{\partial S^2}\right)_{q}&=&\frac{3\pi S+\Lambda S^2-15\eta \pi^2q^2}{16\pi^{3/2}S^{7/2}}.\label{11}
\end{eqnarray}%
It is not difficult to observe that the numerator of $\frac{\partial T}{\partial S}$ coincides with the denominator of $C_q$. One can soon draw the conclusion that the equation $\left(\frac{\partial T}{\partial S}\right)_{q}=0$ has only one positive root for $\eta=-1$. The numerator of $\left(\frac{\partial^2 T}{\partial S^2}\right)_{q}$ turns out to be $2\pi S-12\eta \pi^2q^2$ with the equation $\pi S+\Lambda S^2=3\eta \pi^2q^2$ substituted. $2\pi S-12\eta \pi^2q^2$ is always positive when $\eta=-1$. So the equation $\left(\frac{\partial^2 T}{\partial S^2}\right)_{q}=0$ has no real root. And no $T-S$ criticality exists for the case $\eta=-1$ in the canonical ensemble, which is quite different from the case $\eta=1$ (the RN-AdS black hole).

\begin{figure}
\centerline{\subfigure[]{\label{1a}
\includegraphics[width=8cm,height=6cm]{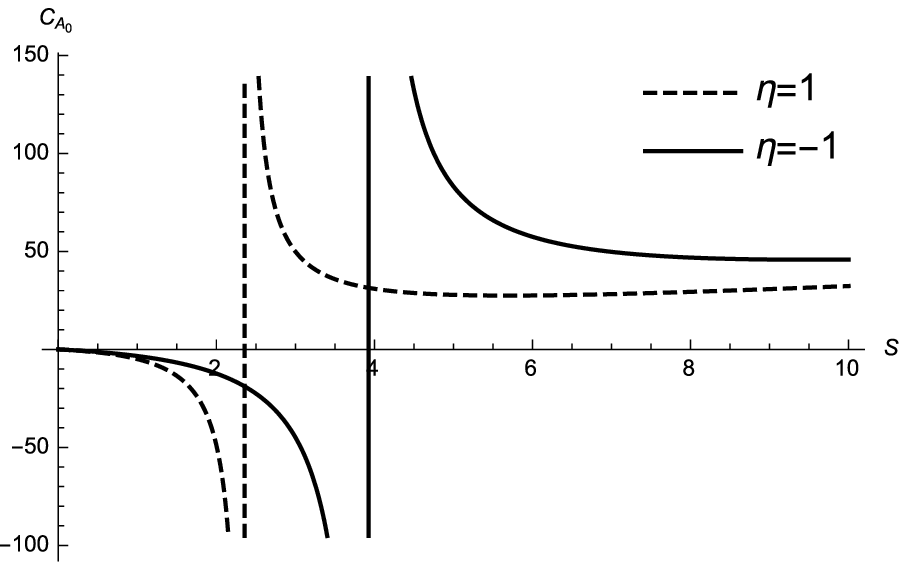}}
\subfigure[]{\label{1b}
\includegraphics[width=8cm,height=6cm]{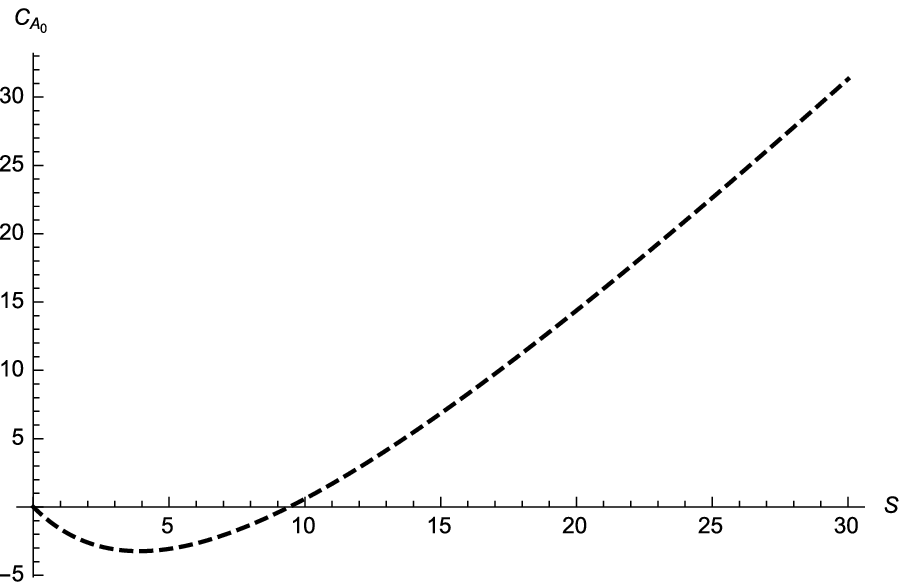}}}
\centerline{\subfigure[]{\label{1c}
\includegraphics[width=8cm,height=6cm]{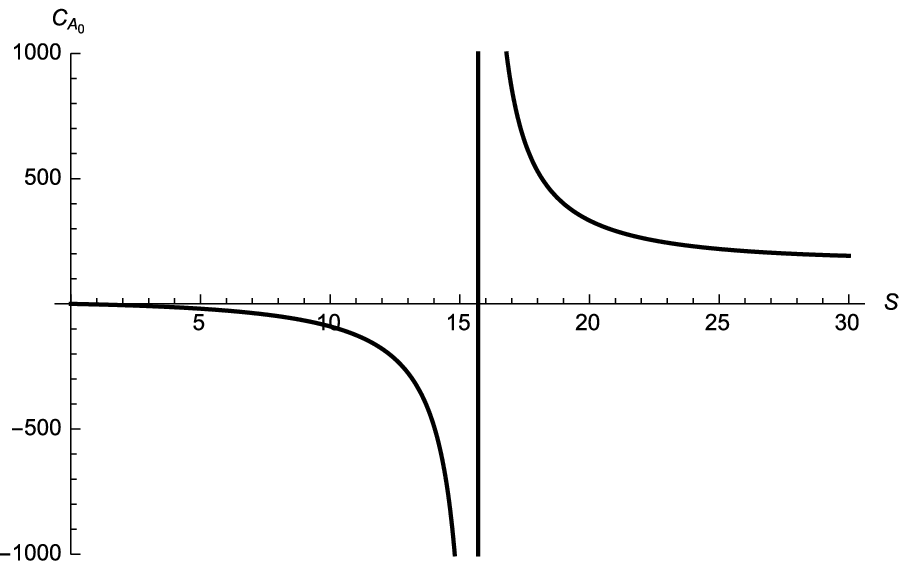}}
\subfigure[]{\label{1d}
\includegraphics[width=8cm,height=6cm]{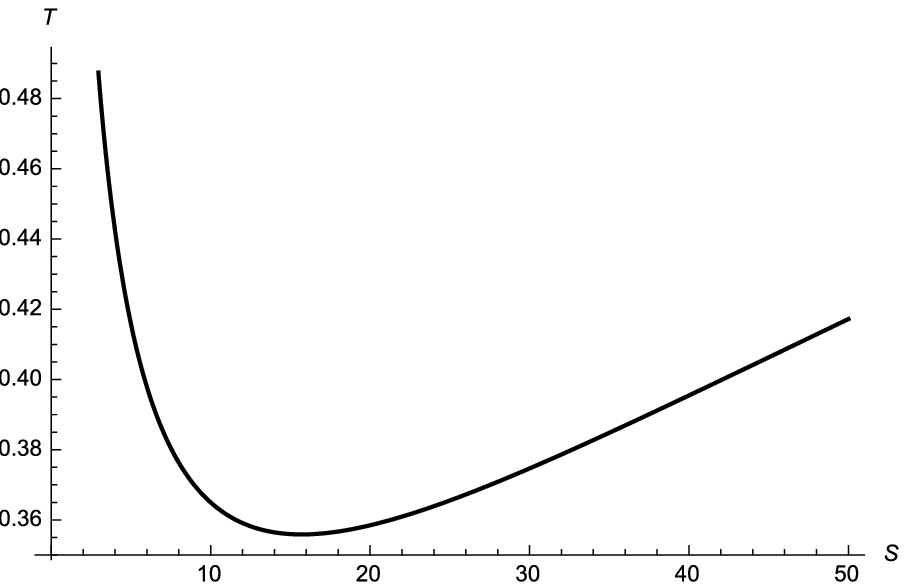}}}
 \caption{Specific heat at constant electric potential for (a) $\eta=1,-1$, $A_0=1/2, \Lambda=-1$ (b) $\eta=1$, $A_0=2, \Lambda=-1$ (c) $\eta=-1$, $A_0=2, \Lambda=-1$ and the Hawking temperature for $\eta=-1$, $A_0=2, \Lambda=-1$ shown in (d)}
\label{fg1}
\end{figure}

For the grand-canonical ensemble, the specific heat at constant electric potential is of interest. Utilizing Eqs. (\ref{4}) and (\ref{6}), one can reexpress the Hawking temperature into the function of $A_0$ and $S$ as
\begin{equation}
T=\frac{\pi-S\Lambda-\eta A_0^2 \pi}{4\pi^{3/2}\sqrt{S}}\;.\label{12}
\end{equation}
Then the specific heat at constant electric potential can be derived as
\begin{equation}
C_{A_0}=T\left(\frac{\partial S}{\partial T}\right)_{A_0}=\frac{2S(-\pi+\Lambda S+\eta \pi A_0^2)}{\pi+\Lambda S-\eta \pi A_0^2} \label{13}\;.
\end{equation}
When $S=\frac{-\pi+\eta A_0^2 \pi}{\Lambda}$, $\pi+\Lambda S-\eta \pi A_0^2=0$. For the case $\eta=1$, this root is physical if the electric potential satisfies $0<A_0<1$. For the case $\eta=-1$, this root is always physical. So the phantom AdS black holes undergo phase transition in the grand-canonical ensemble. To gain an intuitive understanding, the behavior of $C_{A_0}$ is plotted in Fig.\ref{fg1} for comparison. It can be witnessed that for $A_0=1/2, \Lambda=-1$, both the phantom AdS black hole and RN-AdS black holes undergo phase transition although the divergent point of the specific heat differs from each other. However, for $A_0=2, \Lambda=-1$, only phantom AdS black holes undergo phase transition. From the graph of the Hawking temperature, one can see that these choices of parameters lead to positive temperature and the results discussed above make sense physically.

%%%%%%%%%%%%%%%%%%%%%%%%%%%%%%%%%%%%%%%%%%%%%%%%%%%%%%%%%%%%%%%%%%%%%%%%%%%%%
%%%%%%%%%%%%%%%%%%%%%%%%%%%%%%%%%%%%%%%%%%%%%%%%%%%%%%%%%%%%%%%%%%%%%%%%%%%%%%%%

One can also probe the issue of $T-S$ criticality in the grand-canonical ensemble. From Eq. (\ref{12}), one can obtain
\begin{eqnarray}
\left(\frac{\partial T}{\partial S}\right)_{A_0}&=&-\frac{\pi+\Lambda S-\eta \pi A_0^2}{8\pi^{3/2}S^{3/2}},\label{14}
\\
\left(\frac{\partial^2 T}{\partial S^2}\right)_{A_0}&=&\frac{3\pi+\Lambda S-3\eta \pi A_0^2}{16\pi^{3/2}S^{5/2}}.\label{15}
\end{eqnarray}%
Solving the equations $\left(\frac{\partial T}{\partial S}\right)_{A_0}=0$ and $\left(\frac{\partial^2 T}{\partial S^2}\right)_{A_0}=0$, one can obtain the solution $S=0, A_0=\frac{1}{\sqrt{\eta}}$. Obviously, this solution is not physical for neither the case $\eta=1$ nor the case $\eta=-1$. So no critical point can be found for both the RN-AdS black hole and the phantom AdS black hole in the grand-canonical ensemble.

\section{Phase transition of phantom AdS black holes in the extended phase space}
\label{Sec4}

In this section, we would like to probe the phase transition of phantom AdS black holes in the extended phase space.

Firstly, we define the thermodynamic pressure and thermodynamic volume as follows \cite{Kubiznak}
\begin{eqnarray}
P&=&-\frac{\Lambda}{8\pi},\label{16}
\\
V&=&\left(\frac{\partial M}{\partial P}\right),\label{17}
\end{eqnarray}%
where cosmological constant is view as a variable and identified as pressure.

Substituting Eq. (\ref{16}) into Eq. (\ref{3}), one can reexpress the mass into the function of $P$ as
\begin{equation}
M=\frac{r_+}{2}\left(1+\frac{8\pi P}{3}r_+^2+\eta\frac{q^2}{r_+^2}\right)\label{18}\;.
\end{equation}
Then the thermodynamic volume can be calculated as
\begin{equation}
V=\frac{4\pi r_+^3}{3}\;.\label{19}
\end{equation}
The expression above is independent of $\eta$, implying that the phantom AdS black hole shares the same thermodynamic volume with the RN-AdS black hole.

The equation of state can be derived as
\begin{equation}
P=\frac{T}{2r}+\frac{\eta q^2}{8\pi r_+^4}-\frac{1}{8\pi r_+^2}\;.\label{20}
\end{equation}
The difference of phantom AdS black holes and RN-AdS black holes is reflected in the sign of the second term in the equation of state. We will show below just this minor difference leads to completely different results in the existence of $P-V$ criticality.

\begin{figure}
\centerline{\subfigure[]{\label{2a}
\includegraphics[width=8cm,height=6cm]{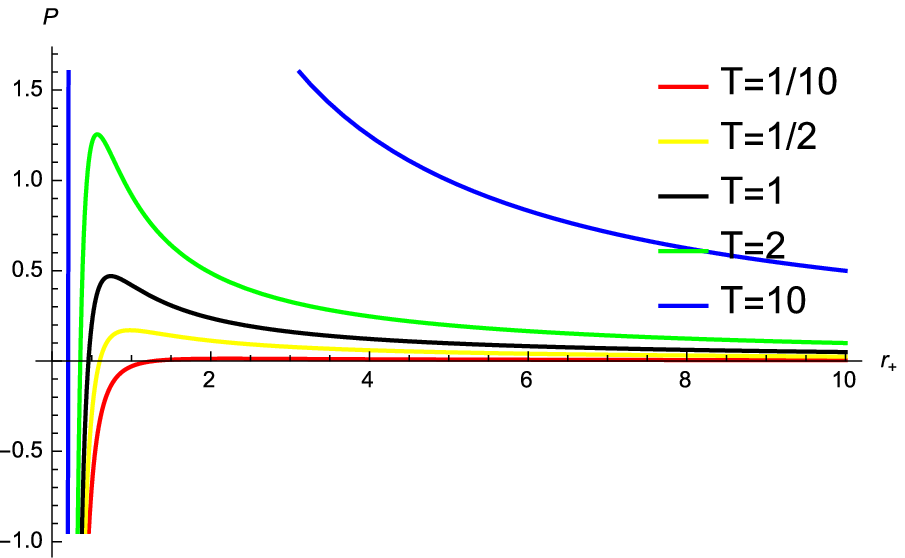}}
\subfigure[]{\label{2b}
\includegraphics[width=8cm,height=6cm]{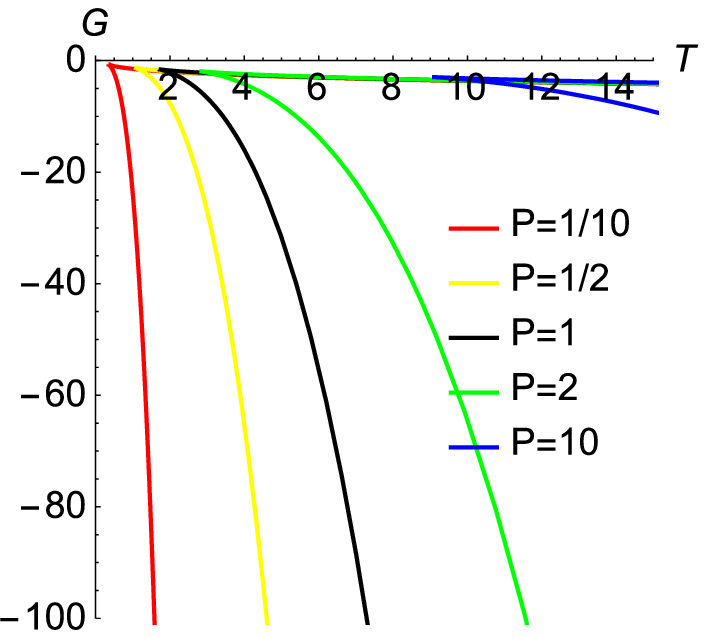}}}
 \caption{(a) The thermodynamic pressure for $\eta=-1$, $q=1$ with various choices of $T$ (b)The Gibbs free energy of phantom AdS black holes for $\eta=-1$, $q=1$ with various choices of $P$}
\label{fg2}
\end{figure}

Based on Eq. (\ref{20}), it is straightforward to derive
\begin{eqnarray}
\left(\frac{\partial P}{\partial r_+}\right)_{T,q}&=&\frac{r_+^2-2\pi r_+^3 T-2\eta q^2}{4\pi r_+^5},\label{21}
\\
\left(\frac{\partial^2 P}{\partial r_+^2}\right)_{T,q}&=&\frac{r_+^2(-3+4\pi r_+ T)+10\eta q^2}{4\pi r_+^{6}}.\label{22}
\end{eqnarray}%
Then the roots of the equations $\left(\frac{\partial P}{\partial r_+}\right)_{T=T_c,r_+=r_c}=0$ and $\left(\frac{\partial^2 P}{\partial r_+^2}\right)_{T=T_c,r_+=r_c}=0$ can be analytically solved as $T_c=\frac{r_c^2-2\eta q^2}{2\pi r_c^3}, r_c=q\sqrt{6\eta}$. Obviously, this solution of $r_c$ is not physical when $\eta=-1$. So there is no $P-V$ criticality for phantom AdS black holes, contrary to RN-AdS black holes. For an intuitive understanding, see Fig.\ref{2a}.

One can further investigate the Gibbs free energy of phantom AdS black holes in the extended phase space, where the mass should be interpreted as enthalpy rather than the internal energy. And the definition of Gibbs free energy reads $G=H-TS=M-TS$. Utilizing Eqs. (\ref{5}), (\ref{18}) and (\ref{20}), one can obtain
 \begin{equation}
G=\frac{r_+}{4}-\frac{2P\pi r_+^3}{3}+\frac{3\eta q^2}{4r_+}\;.\label{23}
\end{equation}
The difference of phantom AdS black holes and RN-AdS black holes lies in the sign of the third term. The behavior of Gibbs free energy of phantom AdS black holes is shown in Fig.\ref{2b}. For each choice of $P$, the curve of the Gibbs free energy can be divided into two branches and the lower branch is more stable due to low free energy. No swallow tail behavior (two stable branches and one unstable branch) can be observed from the Gibbs free energy graph and thus no first order phase transition exists for phantom AdS black holes.

%%%%%%%%%%%%%%%%%%%%%%%%%%%%%%%%%%%%%%%%%%%%%%%%%%%%%%%%%%%%%%%%%%%%%%%%%%%%%
%%%%%%%%%%%%%%%%%%%%%%%%%%%%%%%%%%%%%%%%%%%%%%%%%%%%%%%%%%%%%%%%%%%%%%%%%%%%%%%%

\section{Phantom AdS black holes as heat engines}
\label{Sec5}

\begin{figure}
\centerline{\subfigure[]{\label{3a}
\includegraphics[width=8cm,height=6cm]{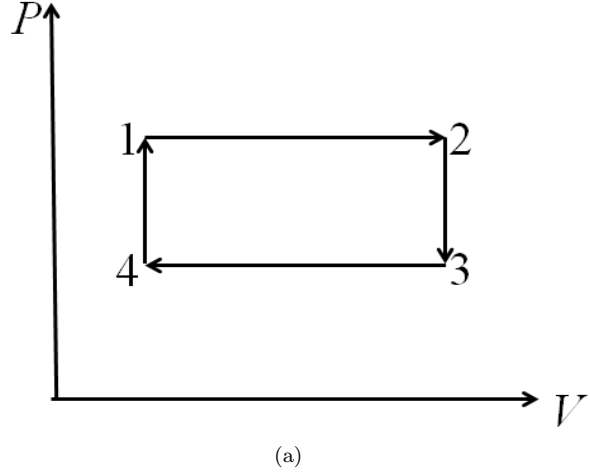}}}
 \caption{The heat engine cycle considered in this paper}
\label{fg3}
\end{figure}

A new kind of heat engine via phantom AdS black holes would be built in this section. To probe its heat engine efficiency, we consider a rectangle cycle consisting of two isochores and two isobars in the $P-V$ plane. See Fig.\ref{fg3} for a sketch of the cycle. Subscripts $1, 2, 3, 4$ are used to denote four corners of the cycle and the physical quantities evaluated at the corresponding corner. From Eqs. (\ref{5}) and (\ref{19}), one can draw the conclusion that the specific heat at constant volume $C_V$ equals to zero. Then no heat flows along the isochores.

%%%%%%%%%%%%%%%%%%%%%%%%%%%%%%%%%%%%%%%%%%%%%%%%%%%%%%%%%%%%%%%%%%%%%%%%%%%%%
%%%%%%%%%%%%%%%%%%%%%%%%%%%%%%%%%%%%%%%%%%%%%%%%%%%%%%%%%%%%%%%%%%%%%%%%%%%%%%%%

Substituting Eq. (\ref{5}) into (\ref{18}), one can further express the mass into the function of $S$
\begin{equation}
M=\frac{3S+8PS^2+3\eta \pi q^2}{6\sqrt{\pi S}}\label{24}\;.
\end{equation}

The heat input $Q_H$ along the isobar $1\rightarrow2$ can be calculated as
\begin{equation}
Q_H=\int^{T_2}_{T_1}C_PdT=\int^{S_2}_{S_1}C_P \left(\frac{\partial T}{\partial S}\right)dS=\int^{S_2}_{S_1}TdS=\int^{H_2}_{H_1}dH=M_2-M_1.\label{25}
\end{equation}%

With Eq. (\ref{24}), $Q_H$ can be obtained as
\begin{eqnarray}
Q_H&=&\frac{1}{2\sqrt{\pi}}(\sqrt{S_2}-\sqrt{S_1})+\frac{4P_1}{3\sqrt{\pi}}(S_2^{3/2}-S_1^{3/2})+\frac{\sqrt{\pi}\eta q^2}{2}(S_2^{-1/2}-S_1^{-1/2}).\label{26}
\end{eqnarray}%

Utilizing Eqs. (\ref{5}) and (\ref{19}), one can obtain the work done along the cycle as
 \begin{equation}
W=(V_2-V_1)(P_1-P_4)=\frac{4(P_1-P_4)(S_2^{3/2}-S_1^{3/2})}{3\sqrt{\pi}}.\label{27}
\end{equation}%

Via Eqs. (\ref{26}) and (\ref{27}), the heat engine efficiency can be derived as
 \begin{equation}
\eta=\frac{W}{Q_H}=\left(1-\frac{P_4}{P_1}\right) \times \frac{4P_1(S_2^{3/2}-S_1^{3/2})}{4P_1(S_2^{3/2}-S_1^{3/2})+\frac{3}{2}(\sqrt{S_2}-\sqrt{S_1})\left(1-\frac{\pi\eta q^2}{\sqrt{S_1S_2}}\right)}.\label{28}
\end{equation}%

From Eqs.(\ref{26}) and (\ref{27}), one can conclude that the existence of phantom field only exerts influence on the heat input $Q_H$, leaving the work unchanged. For the same work done along the cycle, the heat input for the phantom black holes is more than that of RN-AdS black holes. So phantom AdS black holes have a lower heat engine efficiency comparing to RN-AdS black holes. And this may be attributed to the existence of phantom field whose energy density is negative. Moreover, with the increasing of $q$, the heat engine efficiency of phantom AdS black holes decreases while that of RN-AdS black holes increases.

\begin{figure}
\centerline{\subfigure[]{\label{4a}
\includegraphics[width=8cm,height=6cm]{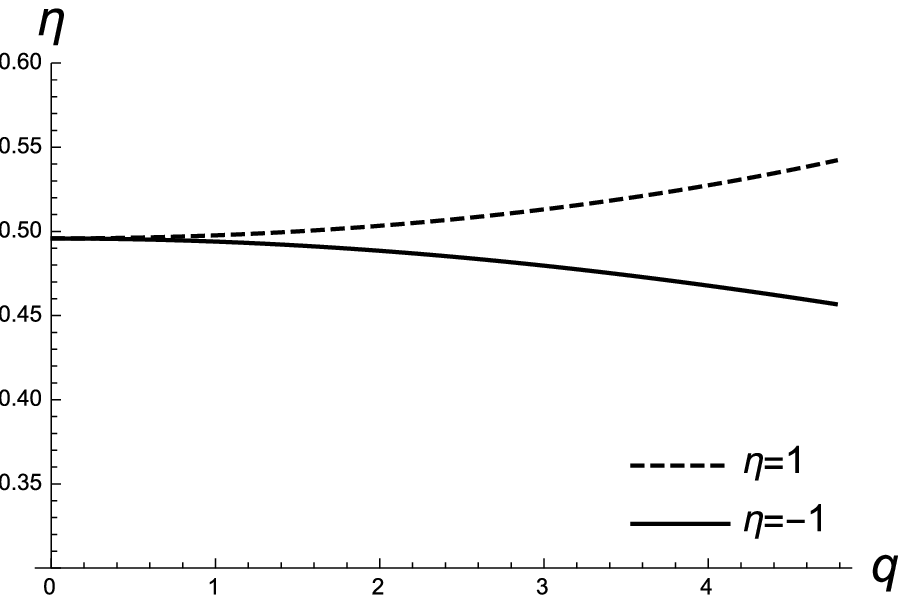}}
\subfigure[]{\label{4b}
\includegraphics[width=8cm,height=6cm]{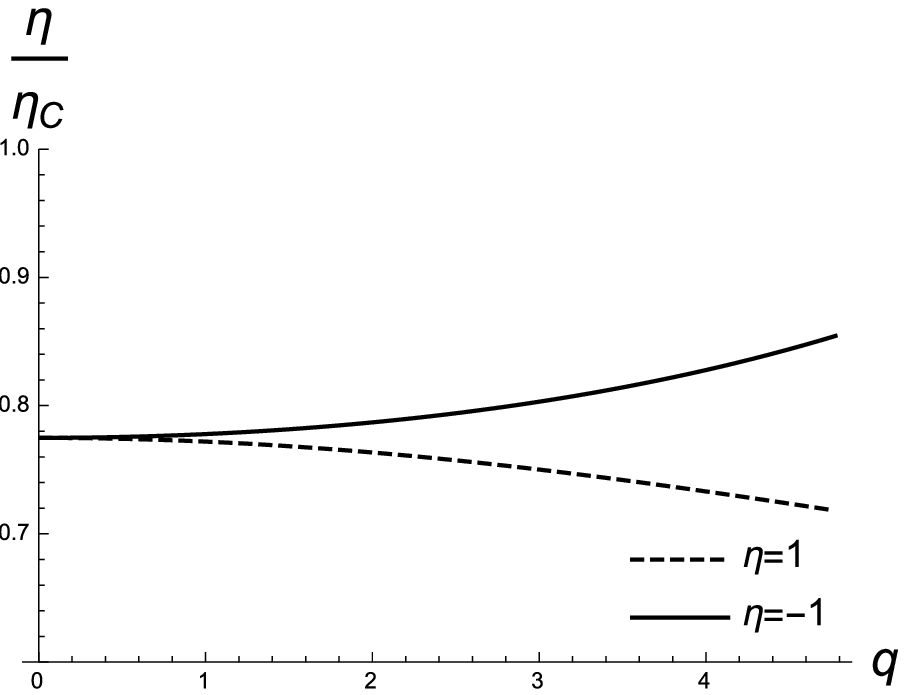}}}
 \caption{A specific example with the parameters chosen as $P_1=2, P_4=1, S_1=5, S_2=10$ (a) The heat engine efficiency $\eta$ vs. $q$ (b) the ratio $\eta/\eta_C$ vs. $q$}
\label{fg4}
\end{figure}

To gain an intuitive understanding, we plot the heat engine efficiency for a specific example $P_1=2, P_4=1, S_1=5, S_2=10$ in Fig. \ref{fg4}. The solid line shows the case $\eta=-1$ while the dashed line shows the case $\eta=1$. Note that the positivity of the Hawking temperature and the mass puts on constraints on the parameter $q$. Considered these constraints (by demanding $T_4>0$ and $M_4>0$), $q$ should not exceed $\sqrt{\frac{215}{3\pi}}$ (for the parameters chosen for the cycle) for the phantom AdS black holes while it should not exceed $\sqrt{\frac{205}{\pi}}$ for RN-AdS black holes. And we have considered these constraints in the graph. It can be witnessed from Fig. \ref{4a} that the heat engine efficiency of phantom AdS black holes decreases with $q$ while the heat engine efficiency of RN-AdS black holes increases with $q$. And this observation is well in accord with the theoretical derivation above.

One can also compare the heat engine efficiency with the well-known Carnot efficiency $\eta_C$ and investigate the ratio $\eta/\eta_C$. The highest temperature of the heat engine cycle $T_H$ should be $T_2$ and the lowest temperature of the heat engine cycle $T_C$ should be $T_4$ for our cycle. So 
 \begin{equation}
 \eta_C=1-\frac{T_C}{T_H}=1-\frac{S_2^{3/2}(S_1+8P_4S_1^2-\eta\pi q^2)}{S_1^{3/2}(S_2+8P_1S_2^2-\eta\pi q^2)}.\label{29}
\end{equation}%
For the case $\eta=-1$, both the numerator and denominator of the second term in the right hand side of the above expression increase. To see the combined effect, one can compare the value of $\eta_C$ for both cases.
 \begin{equation}
 \eta_C|_{\eta=1}- \eta_C|_{\eta=-1}=\frac{2\pi q^2S_2^{3/2}(S_2-S_1+8P_1S_2^2-8P_4S_1^2)}{S_1^{3/2}(-\pi q^2+S_2+8P_1S_2^2)(\pi q^2+S_2+8P_1S_2^2)}>0.\label{29}
\end{equation}%
So the carnot efficiency $\eta_C$ of phantom AdS black holes is lower than that of RN-AdS black holes. Fig. \ref{4b} shows the behavior of ratio $\eta/\eta_C$ for the case $P_1=2, P_4=1, S_1=5, S_2=10$. One can see that the ratio increases with $q$ for the case $\eta=-1$ while it decreases with $q$ for the case $\eta=1$. So the existence of phantom field increases the possibility of approaching the Carnot limit, which is of interest to researchers.

%%%%%%%%%%%%%%%%%%%%%%%%%%%%%%%%%%%%%%%%%%%%%%%%%%%%%%%%%%%%%%%%%%%%%%%%%%%%%
%%%%%%%%%%%%%%%%%%%%%%%%%%%%%%%%%%%%%%%%%%%%%%%%%%%%%%%%%%%%%%%%%%%%%%%%%%%%%%%%

\section{Conclusions}
\label{Sec6}
  We study the phase transition and the heat engine efficiency of phantom AdS black holes in this paper. Firstly, we consider the phase transition in the non-extended phase space. Except the specific heat at constant charge which has already been studied in the former literature, we probe the possibility of $T-S$ criticality in both the canonical ensemble and grand-canonical ensemble. In the canonical ensemble, it is shown that no $T-S$ criticality exists for the case $\eta=-1$ (the phantom AdS black hole), which is quite different from the case $\eta=1$ (the RN-AdS black hole). Contrary to the canonical ensemble, no critical point can be found for neither phantom AdS black holes nor RN-AdS black hole in the grand-canonical ensemble. Moreover, we study the specific heat at constant electric potential $C_{A_0}$. It is shown that when the electric potential satisfies $0<A_0<1$, both phantom AdS black holes and RN-AdS black holes undergo phase transition in the grand-canonical ensemble. However, when the electric potential satisfies $A_0>1$, only phantom AdS black holes undergo phase transition.

Secondly, we investigate the phantom AdS black holes in the extended phase space. We identify the cosmological constant as thermodynamic pressure and define its conjugate quantity as the thermodynamic volume. Then we show that there is no $P-V$ criticality for phantom AdS black holes. This observation is contrary to RN-AdS black holes. We further study the behavior of Gibbs free energy of phantom AdS black holes. It can be witnessed that the curve of the Gibbs free energy can be divided into two branches and the lower branch is more stable due to low Gibbs free energy. No swallow tail behavior can be observed from the Gibbs free energy graph and thus no first order phase transition exists for phantom AdS black holes.

Thirdly, we define a new kind of heat engine via phantom AdS black holes. It is shown that the existence of phantom field only exerts influence on the heat input $Q_H$, leaving the work unchanged. For the same work done along the cycle, the heat input for the phantom black holes is more than that of RN-AdS black holes. So phantom AdS black holes have a lower heat engine efficiency comparing to RN-AdS black holes. And this may be attributed to the existence of phantom field whose energy density is negative. Moreover, with the increasing of $q$, the heat engine efficiency of phantom AdS black holes decreases while that of RN-AdS black holes increases. Moreover, we compare the heat engine efficiency with the well-known Carnot efficiency $\eta_C$ and investigate the ratio $\eta/\eta_C$. We show that this ratio increases with $q$ for phantom AdS black holes while it decreases with $q$ for RN-AdS black holes. So the existence of phantom field increases the possibility of approaching the Carnot limit, which is of interest to researchers.

 \section*{Acknowledgements}

 This research is supported by National Natural Science Foundation of China (Grant No.11605082), and in part supported by Natural Science Foundation of Guangdong Province, China (Grant Nos.2016A030310363, 2016A030307051, 2015A030313789).

\end{document}